\documentclass[12pt,a4paper]{article}
\pdfoutput=1
\usepackage[T1]{fontenc} 
\usepackage{amsmath}
\usepackage{amsfonts}
\usepackage{amssymb}
\usepackage{graphicx}
\usepackage{subfig}
\usepackage{hyperref}
\usepackage{color}

\newcommand{\HH}{{\cal H}}

\newcommand{\be}{\begin{equation}}
\newcommand{\ee}{\end{equation}}
\newcommand{\beq}{\begin{equation}}
\newcommand{\eeq}{\end{equation}}
\newcommand{\bea}{\begin{eqnarray}}
\newcommand{\eea}{\end{eqnarray}}
\def\be{\begin{equation}}
\def\ee{\end{equation}}

\def\beq{\begin{equation}}
\def\eeq{\end{equation}}

\newcommand{\rmd}{\textrm{d}}

\newcommand{\dA}{\delta^{(A)}}
\newcommand{\dB}{\delta^{(B)}}
\newcommand{\tA}{\theta^{(A)}}
\newcommand{\tB}{\theta^{(B)}}

\begin{document}
\def\thefootnote{\fnsymbol{footnote}}

\begin{center}
\Large{\textbf{Single-Field Consistency Relations of Large Scale Structure \\Part III: Test of the Equivalence Principle}} \\[0.5cm]
\end{center}
\vspace{0.5cm}

\begin{center}

\large{Paolo Creminelli$^{\rm a}$, J\'er\^ome Gleyzes$^{\rm b,\rm c}$, Lam Hui$^{\rm d}$, \\Marko Simonovi\'c$^{\rm e, \rm f}$ and
 Filippo Vernizzi$^{\rm b}$}
\\[0.5cm]

\small{
\textit{$^{\rm a}$ Abdus Salam International Centre for Theoretical Physics\\ Strada Costiera 11, 34151, Trieste, Italy}}

\vspace{.2cm}

\small{
\textit{$^{\rm b}$ CEA, Institut de Physique Th\'eorique, F-91191 Gif-sur-Yvette c\'edex, France}}

%\vspace{.2cm}
\small{
\textit{CNRS, Unit\'e de recherche associ\'ee-2306, F-91191 Gif-sur-Yvette c\'edex, France}}
\vspace{.2cm}

\small{
\textit{$^{\rm c}$ Universit\'e Paris Sud, 15 rue George Cl\'emenceau, 91405,  Orsay, France}}

\vspace{.2cm}

\small{
\textit{$^{\rm d}$ Physics Department and Institute for Strings, Cosmology and Astroparticle Physics, Columbia University, New York, NY 10027}}

\vspace{.2cm}

\small{
\textit{$^{\rm e}$ SISSA, via Bonomea 265, 34136, Trieste, Italy}}

\vspace{.2cm}

\small{
\textit{$^{\rm f}$ Istituto Nazionale di Fisica Nucleare, Sezione di Trieste, via Bonomea 265, 34136, Trieste, Italy}}

\vspace{.2cm}

\end{center}

\vspace{.8cm}

\hrule \vspace{0.3cm}
\noindent \small{\textbf{Abstract}\\ 
The recently derived consistency relations for Large Scale Structure do not hold if the Equivalence Principle (EP) is violated. We show it explicitly in a toy model with two fluids, one of which is coupled to a fifth force. We explore the constraints that galaxy surveys can set on EP violation looking at the squeezed limit of the 3-point function involving two populations of objects. We find that one can explore EP violations of order $10^{-3} \div 10^{-4}$ on cosmological scales. Chameleon models are already very constrained by the requirement of  screening within the Solar System and only a very tiny region of the parameter space can be explored with this method. We show that no violation of the consistency relations is expected in Galileon models.}
\\
\noindent
\hrule
\def\thefootnote{\arabic{footnote}}
\setcounter{footnote}{0}

\newpage

\section{Introduction}
Experimental tests of the Equivalence Principle (EP) are, from Galileo to modern torsion balance experiments, a prototypical example of the scientific method. The impressive modern limits on the equality of inertial and gravitational mass testify that we understand gravity very well, at least on scales much shorter than the Hubble size. On the other hand, the observed acceleration of the Universe may suggest that something new happens to gravity at very large distances. It is at first difficult to imagine how to test the EP on scales comparable to the size of the Universe, since even the most patient experimentalist cannot follow the fall of astrophysical objects for lengths and timescales comparable to Hubble. In this paper we show that this kind of test is indeed possible:  we do not have to {\em wait} for things to fall, we just have to look at their {\em final position}, provided we make the correct guess about their initial conditions long back in time. It is like saying that Galileo could have simply studied the arrival time of the different rolling balls along the inclined plane, provided somebody had told him in advance the initial conditions at the top of the plane. Usually, initial conditions are part of the experimental setup and not something that can be predicted from the theory, or at least this was the situation for Galileo. Nowadays we think we know the initial conditions of our Universe, at least in a statistical, if not deterministic, sense. All the experiments are compatible with the simple picture of Gaussian initial conditions and this is what we are going to assume throughout this paper, keeping of course in mind that a deviation from this assumption would be a big discovery on its own\footnote{What we need to assume is the absence of non-Gaussianity in the squeezed limit.}. The absence of non-Gaussianity, i.e.~the statistical independence of the Fourier modes, tells us that the homogeneous gravitational field where the experiment will take place does not affect the (statistics of) initial conditions for the small objects whose fall we test. This educated assumption about initial conditions allows to test the EP on cosmological scales by 
simply measuring the position of different astrophysical objects at a given time. 

This paper is a natural continuation of \cite{Creminelli:2013mca,Creminelli:2013poa}, where we showed, following \cite{Kehagias:2013yd,Peloso:2013zw}, that for single-field inflationary models and if the EP holds, certain consistency relations for cosmological correlation functions can be derived. The violation of the consistency relations in modified gravity theories has been recently discussed in \cite{Peloso:2013spa,Kehagias:2013rpa,Valageas:2013cma,HHX2013} (with some differences that we are going to point out). In this paper we will concentrate on equal-time correlators, which are the most relevant observationally, and on the 3-point function which, in the non-relativistic limit, reads
\be
\lim_{q\to 0} \langle \delta_{\vec q}(\eta) \delta^{(A)}_{\vec k_1}(\eta) \delta^{(B)}_{\vec k_2}(\eta) \rangle' = \bigg( \epsilon \, \frac{\vec{k}\cdot\vec{q}}{q^2}  + {\cal{O}}\big[ (q/k)^0 \big] \bigg) P(q,\eta) P_{AB}(k,\eta)\, .\label{VEP}
\ee
The notation requires some explanation. A prime on the correlation
function on the left-hand side  indicates that the momentum conserving
Dirac function has been removed. $\delta^{(A)}$ and $\delta^{(B)}$ are
the number densities of the two classes of objects (e.g.~galaxies with
different mass) we want to compare in their fall and $P_{AB}(k,\eta)$
their cross power spectrum, with $\vec k \equiv (\vec k_1 - \vec k_2
)/2$. The third mode $\delta_{\vec q}$ with small momentum $q$
corresponds to the approximately homogeneous gravitational field where
objects $A$ and $B$ fall. It is treated in the linear regime and can
be measured using any probe we like. If objects $A$ and $B$ fall in
the same way, then $\epsilon$ vanishes. Conversely, as we will see, a
deviation from the EP for the two classes of objects induces a
non-zero $\epsilon$. Equation \eqref{VEP} represents a violation of the
consistency relation, which tells us that there should
be no $k/q$ term in such an equal-time correlator, if the EP and single-field initial condition were
respected. The actual size of the violation of the consistency relation
is model-dependent.
In Section \ref{sec:modelAB} we are going to calculate $\epsilon$ in a simple model in which the objects $A$ and $B$ have a different coupling with a long range fifth force. Although modified gravity models can be significantly more complicated, this will represent our benchmark model. In Section \ref{sec:SN} we are going to study the limits one will be able to put on the parameter $\epsilon$ in future surveys with a simple estimate of the cumulative signal-to-noise in the bispectrum. Notice that exchanging $A$ with $B$ is equivalent to flipping the sign of $\vec q$ so that the relation for $A=B$ trivially vanishes: only for two different kinds of objects the consistency relation can be violated\footnote{This point has not been made explicit in Ref.~\cite{Kehagias:2013rpa}, where they concentrate on correlation functions for the same class of objects.}. 
% Divergent!

What kind of models are expected to violate the EP on cosmological scales? One possibility is the existence of some non-universal long-range force, another is the EP violation induced on macroscopic objects by one of the screening mechanisms, which hide the deviations from GR on short scales, where stringent experimental bounds apply. We will review these possibilities in Sec.~\ref{sec:models}. It is fair to anticipate that most of these models give a negligible signal for our test, either because of other experimental constraints or because the EP violation is anyway suppressed. We think, however, that this does not diminish the interest in testing the EP on cosmological scales. Indeed one has to admit that none of the models which modify gravity on large scales addresses the cosmological constant problem, which is the main reason why we are interested in modifications of gravity in the first place. Therefore, if gravity changes on large scales in a way connected with the cosmological constant, we expect something much more dramatic and interesting than the theories studied so far. From this point of view, a test of the basic tenet of GR on cosmological scales is surely worthwhile. 

One can read eq.~\eqref{VEP}, when the EP holds, i.e.~$\epsilon = 0$, as the statement that there is no velocity bias between species $A$ and $B$ on large scales: the long mode induces exactly the same velocity for all objects. 
It is important to stress that this holds even considering {\em
  statistical} velocity bias. Objects do not form randomly, but in
special places of the density field: therefore, even if they locally
fall together with the dark matter, there can be a velocity bias in a
statistical sense \cite{Desjacques:2008jj,Desjacques:2009kt}. However,
the arguments of
\cite{Creminelli:2013mca,Creminelli:2013poa,Kehagias:2013yd} tell us
that the long mode (at leading order in $q$) is equivalent to a change
of coordinates. Apart from this change of coordinates the long mode
affects neither the dynamics nor the {\em statistics} of short
modes. Therefore, the EP implies that the statistical velocity bias
disappears on large scales: again, this statement is completely
non-perturbative in the short scales and includes the effect of
baryons. For the case of dark matter only, we know that the
statistical velocity bias vanishes on large scales as $\sim q^2 R^2$,
where $R$ is a length scale of order the Lagrangian size of the
object; this can be calculated by looking at the statistics of peaks
\cite{Desjacques:2008jj,Desjacques:2009kt} and verified in numerical
simulations \cite{Elia:2011ds}. One expects that the effect of
statistical velocity bias is therefore subdominant with respect to the
unknown corrections in eq.~\eqref{VEP} since $k \lesssim
R^{-1}$.(\footnote{
Reference~\cite{Peloso:2013spa} 
    quotes from \cite{Elia:2011ds}
  that, for $q = 0.05 h$Mpc$^{-1}$, objects in the range $(25 \div 40)
  \cdot 10^{12} h^{-1} M_\odot$ have a velocity bias of $1.05$
  compared to dark matter. This effect is more important than the unknown
  corrections ${\cal O}[ (q/k)^{0} ]$ to the consistency relation
  \eqref{VEP} only for $k \gtrsim R^{-1}/\sqrt{0.05}$, where
$R$ is the Lagrangian size of the objects. However, it is difficult to
measure the correlation function of objects on scales smaller than their
Lagrangian size.})

\section{\label{sec:modelAB}An example of equivalence principle violation}

In this section we are going to study a toy model in which the Universe is composed of two non-relativistic fluids $A$ and $B$, 
with the latter coupled to a scalar field mediating a fifth force. For
example, the two fluids could be baryons and dark matter but, with some modifications that we will discuss below, the model can also describe two populations of astrophysical objects, say different types of galaxies.
If the scalar field $\varphi$ has a negligible time evolution, the continuity equations of the two fluids are the same, 
\be
\delta_X' + \vec \nabla \cdot [ (1+ \delta_X ) \vec v_X] = 0\;, \quad X=A,B\;, \label{continuity}
\ee
where a prime denotes the derivative with respect to the conformal time  $\eta \equiv \int dt/a(t)$ (we assume a flat FRW metric, with scale factor $a$), ${}' \equiv \partial_\eta$. The Euler equation of $B$ contains the fifth force, whose coupling is parameterized by $\alpha$,
\begin{align}
\vec v_A' + \HH \vec v_A + (\vec v_A \cdot \vec \nabla) \, \vec v_A &= - \vec \nabla \Phi \;, \label{euler_A} \\ 
\vec v_B' + \HH \vec v_B  + (\vec v_B \cdot \vec \nabla) \, \vec v_B &= - \vec \nabla \Phi - \alpha \vec \nabla \varphi\;, \label{euler_B}
\end{align}
where $\HH \equiv \partial_\eta a /a$  is the comoving Hubble parameter.
To close this system of equations we need Poisson's equation and the evolution equation of the scalar field. Assuming that the scalar field stress-energy tensor is negligible, only matter appears as a source in the Poisson's equation,
\be
\nabla^2 \Phi = 4 \pi G \, \rho_{\rm m} \, \delta = 4 \pi G \, \rho_{\rm m} \, ( w_A \delta_A + w_B \delta_B)\;, \label{poisson}
\ee
where $\rho_{\rm m}$ is the total matter density and  $w_X\equiv \rho_X/\rho_{\rm m}$ is the density fraction of the $X$ species.
Moreover, in the non-relativistic approximation we can neglect time derivatives in comparison with spatial gradients and the equation for the scalar field reads
\be
\nabla^2  \varphi =  \alpha \cdot 8 \pi G \rho_{\rm m} w_B \delta_B \;, \label{varphi}
\ee
where we have neglected the mass of the scalar field, assuming we are on scales much shorter than its Compton wavelength.

Let us start with the linear theory and, following \cite{Saracco:2009df}, look for two of the four independent solutions of the system in which the density and the velocity of the species $B$ differ from those of the species $A$ by a (possibly time-dependent) bias factor $b$,
\begin{align}
\dA_{\vec k}(\eta) &= D (\eta)\, \delta_0(\vec k) \;, \label{deltaA}\\ 
\tA_{\vec k}(\eta) &= - \HH (\eta) f (\eta) \dA_{\vec k}(\eta) \;, \label{thetaA} \\ 
\dB_{\vec k}(\eta) &= b (\eta)\dA_{\vec k}(\eta)\;, \label{deltaB} \\ 
\tB_{\vec k}(\eta) &= - \HH (\eta)  f(\eta) \dB_{\vec k}(\eta)\;, \label{thetaB}
\end{align}
where  we have defined $\theta^{(X)} \equiv \vec \nabla \cdot \vec v_X$ and $\delta_0(\vec k)$ is a Gaussian random variable. 
Plugging this ansatz in eqs.~\eqref{continuity}--\eqref{varphi} and using the background Friedmann equations for a flat universe, we find, at linear order,
\begin{align}
&f=\frac{\rmd\ln D}{\rmd \ln a} \;, \label{fdef}\\
& \frac{\rmd f}{\rmd  \ln a} + f^2 +\left(2- \frac32 \Omega_{\rm m}\right)f  -\frac32 \Omega_{\rm m} (w_A + w_B b) =0 \;, \label{fevol}\\
& \frac{\rmd  b}{\rmd \ln a} =0 \;, \label{bevol}\\
&w_B b + w_A \left(1-\frac1b\right) - w_B (1+2\alpha^2)=0 \label{balpharelation}\;.
\end{align}
Using eqs.~\eqref{fdef} and \eqref{fevol}, the linear growth factor $D$ satisfies a second-order equation,
\be
\frac{\rmd ^2 D}{\rmd  \ln a^2} + \left( 2- \frac32 \Omega_{\rm m} \right) \frac{\rmd  D}{\rmd  \ln a} - \frac32 \Omega_{\rm m} (w_A+w_B b) D =0\;, \label{Devol}
\ee
whose growing and decaying solutions are $D_+$ and $D_-$. Note that eq.~\eqref{bevol} implies that the bias $b$ is time independent. In the absence of EP violation ($\alpha = 0$) we get $b=1$ (using $w_A+w_B=1$) and we recover from eq.~\eqref{Devol} the usual evolution of the growth of matter perturbations.

Following \cite{Crocce:2005xy,Somogyi:2009mh}, we introduce $y \equiv \ln D_+$ as the time variable.
Defining the field multiplet
\be 
\Psi_a \equiv \left( \begin{array}{c}
\dA \\
-\tA/\mathcal{H}f_+ \\
\dB \\
-\tB/\mathcal{H}f_+ \end{array} \right)\;,
\ee
the equations of motion of the two fluids can be then written in a very compact form as
\be
\partial_y \Psi_a (\vec k) + \Omega_{ab} \Psi_b (\vec k) = \gamma_{abc} \Psi_b (\vec k_1) \Psi_c (\vec k_2)\;, \label{EOM}
\ee
where integration over $\vec k_1$ and $\vec k_2$ is implied on the right-hand side. The entries of $\gamma_{abc}$ vanish except for
\be
\label{kernels}
\begin{split}
\gamma_{121} &= \gamma_{343} = (2\pi)^3\delta_D(\vec k - \vec k_1 - \vec k_2) \, \frac{\vec k_1 \cdot (\vec k_1 + \vec k_2)}{k_1^2} \;, \\
\gamma_{222} &= \gamma_{444} = (2\pi)^3\delta_D(\vec k - \vec k_1 - \vec k_2) \, \frac{\vec k_1 \cdot \vec k_2 (\vec k_1 + \vec k_2)^2}{2 k_1^2 k_2^2} \;,
\end{split}
\ee
the matrix $\Omega_{ab}$ reads 
\be
\label{Omega_ab}
 \Omega_{ab}= \left( \begin{array}{cccc}
0 &-1&0&0 \\
-\frac32 \frac{\Omega_{\rm m}}{f_+^2} w_A & \frac32\frac{\Omega_{\rm m}}{f_+^2}(w_A+b w_B) -1&-\frac32 \frac{\Omega_{\rm m}}{f_+^2} w_B&0\\
0&0&0&-1 \\
-\frac32 \frac{\Omega_{\rm m}}{f_+^2} w_A &0&-\frac32 \frac{\Omega_{\rm m}}{f_+^2}( w_B b +w_A\left(1-\frac1b\right))& \frac32 \frac{\Omega_{\rm m}}{f_+^2}(w_A+bw_B)-1\end{array} \right)\, ,
\ee
and we have employed eq.~\eqref{balpharelation} to replace the dependence on $\alpha^2$ by a dependence on the bias $b$.
The solution of eq.~\eqref{EOM} can be formally written as 
\be
\label{solution}
\Psi_a(y) = g_{ab}(y) \phi_b + \int^y_0 \rmd y' g_{ab} (y-y') \gamma_{bcd} \Psi_c(y') \Psi_d(y') \;,
\ee
where $\phi_b$ is the initial condition, $\phi_b = \Psi_b(y=0)$, and $g_{ab}(y)$ is the linear propagator which is given by \cite{Crocce:2005xy}
\be
g_{ab}(y) = \frac{1}{2 \pi i} \int^{\xi+i\infty}_{\xi-i \infty} \rmd \omega\; (\omega I + \Omega )^{-1}_{ab} e^{\omega y}\;,
\ee
where $\xi$ is a real number larger than the real parts of the poles of $(\omega I + \Omega)^{-1}$.

In the following we consider small couplings to the fifth force, $\alpha^2 \ll 1$, which by virtue of eq.~\eqref{balpharelation} implies $b \simeq 1 $. In this case, it is reasonable to use the approximation $f_+^2 \simeq \Omega_{\rm m}$, which for $b=1$ is very good  throughout the whole evolution \cite{Scoccimarro:1997st}. We choose to use this approximation because it considerably simplifies the presentation but one can easily drop it and make an exact computation.

The linear evolution is characterized by four modes. Expanding for small $b-1$, apart from the ``adiabatic'' growing and decaying modes already introduced above, respectively going as $D_+ = e^y$ and $D_- = e^{-\frac32 [1+ w_B (b-1)] y  }$, one finds two ``isodensity'' modes, one decaying as $D_{\rm i} =  e^{-\frac12 [1+ 3(1+w_A) (b-1)] y}$ and an almost constant one going as $D_{\rm c}= e^{3w_A (b-1)  \alpha^2 y}$.(\footnote{With an abuse of language, we  denote the modes $(+)$ and $(-)$ as adiabatic and $({\rm i})$ and $({\rm c})$ as isodensity even though, strictly speaking, they do not correspond to the usual notion of adiabatic and isocurvature. Indeed, $(+)$ and $(-)$  correspond to $\delta_A = \delta_B/b$  and not to $\delta_A = \delta_B$ as in the standard adiabatic case without a fifth force, while $({\rm i})$ and $({\rm c})$ yield $w_A \dA + b w_B \dB = 0$ instead of $w_A \dA + w_B \dB = 0$ which one finds in the standard isodensity case (see \cite{Bernardeau:2011vy} for a discussion of adiabatic and isodensity modes in the standard case $b=1$). })

We are interested in the equal-time 3-point function involving the two species. In particular, we compute 
\be
\langle \delta_{\vec k_3} (\eta) \dA_{\vec k_1}(\eta) \dB_{\vec k_2}(\eta) \rangle = w_{A}  \langle \Psi_1(k_3,\eta) \Psi_1(k_1,\eta) \Psi_3 (k_2,\eta)\rangle +w_{B}  \langle \Psi_3(k_3,\eta) \Psi_1(k_1,\eta) \Psi_3 (k_2,\eta) \rangle  \; ,
\label{3-p}
\ee 
where $\delta \equiv w_A \dA + w_B \dB$. The calculation can be straightforwardly done at tree level by perturbatively expanding the solution \eqref{solution} as $\Psi_a = \Psi^{(1)}_a + \Psi^{(2)}_a+ \ldots$, which up to second order in $\delta_0$ yields 
\be
\begin{split}
\Psi^{(1)}_a(y) &= g_{ab}(y) \phi_b \;, \\
\Psi^{(2)}_a(y) &= \int^y_0 \rmd y' g_{ab} (y-y') \gamma_{bcd} \Psi^{(1)}_c(y') \Psi^{(1)}_d(y') \;,
\end{split}
\ee
and by applying Wick's theorem over the Gaussian initial conditions. In the squeezed limit, the expression for \eqref{3-p} simplifies considerably.
Assuming that the initial conditions are in the most growing mode, i.e.~they are given by $\phi_a (\vec k) = u_a \delta_0 (\vec k)$ with $u_a = (1,1,b,b)$, at leading order in $b-1$ 
one finds 
\be
\begin{split}
\lim_{q\to 0}\langle \delta_{\vec q} (\eta) \dA_{\vec k_1}(\eta) \dB_{\vec k_2}(\eta) \rangle' &\simeq -(b-1)P(q,\eta)P_{AB}(k,0)\frac{\vec k \cdot \vec q }{q^2}  \\
&\times \int^y_0 \rmd y'  e^{2y'}\big[ g_{11} (y-y') +g_{12}(y-y')  - g_{31}(y-y') - g_{32}(y-y') \big]  \;,
\end{split}
\ee
which shows that the long wavelength adiabatic evolution   has no effect on the 3-point function\footnote{For $b=1$ one finds $g^{(+)}_{11} = g^{(+)}_{31}$, $g^{(+)}_{12} = g^{(+)}_{32}$, $g^{(-)}_{11} = g^{(-)}_{31}$ and $g^{(-)}_{12} = g^{(-)}_{32}$.} \cite{Bernardeau:2011vy,Bernardeau:2012aq}. 
As before, the prime on the correlation function denotes that the delta function of momentum conservation   has been dropped.
Retaining the  most growing contribution and using $b \simeq 1 + 2 w_B \alpha^2$ one finally finds
\be
\label{eq:EPviolation}
\lim_{q\to 0} \langle \delta_{\vec q}(\eta) \delta^{(A)}_{\vec k_1}(\eta) \delta^{(B)}_{\vec k_2}(\eta) \rangle' \simeq \frac75 w_B \, \alpha^2 \, \frac{\vec{k}\cdot\vec{q}}{q^2} P(q,\eta) P_{AB}(k,\eta) \, .
\ee

	The result that we obtained remains qualitatively the same if
        $A$ or $B$ represent extended objects, for example a fluid of
        galaxies of a given kind \cite{Chan:2012jj}. For instance, one
        can take $A$ to represent galaxies that are not coupled to the
        fifth force because they are screened (see
        Sec.~\ref{sec:models}), while $B$ represents the dark matter
        fluid. In this case one should start from initial conditions
        in which there is a bias between the galaxy and the dark
        matter overdensities: $u_a=(b_g,1,b,b)$.   This is equivalent
        to exciting decaying modes, given that asymptotically the
        galaxy bias becomes unity ($b_g$ is the initial galaxy bias).
Consequently, the result \eqref{eq:EPviolation} will be different.
Still, it is straightforward to check that, as expected, there is no
$1/q$ divergence if the EP is not violated, i.e.~$\alpha = 0$. In the
limit $w_A \ll 1$ (i.e. the screened galaxies contribute a subdominant
component of the overall mass density) and keeping only the slowest decaying mode, one gets (in this case we take the long mode to be dark matter only)
\be
\label{eq:EPviolationbias}
\lim_{q\to 0} \langle \delta^{(B)}_{\vec q}(\eta) \delta^{(A)}_{\vec k_1}(\eta) \delta^{(B)}_{\vec k_2}(\eta) \rangle' \simeq \frac7{5}\, \alpha^2 \, \frac{\vec{k}\cdot\vec{q}}{q^2}  \left(1+\frac{10}7(b_g-1)e^{-(y-y_0)}\right) P(q,\eta) P_{AB}(k,\eta)\, .
\ee
Notice that $y_0$ here represents the initial value when the local galaxy bias $b_g$ is set up.
	
	Another complication in this case comes from the fact that objects become screened only at a certain stage of their evolution, so that the coupling of the fluid $A$ with the scalar is time-dependent. All this modifies the numerical value on the right-hand side of eq.~\eqref{eq:EPviolation}. In any case, given the model-dependence of the result, we stick to eq.~\eqref{eq:EPviolation} as our benchmark model when discussing the capabilities of experiments to constrain EP violation.

\section{\label{sec:SN} Detecting an equivalence principle violation}

In this section we want to explore how well we can constrain the violation of the EP in our toy model using large scale structure surveys. We will use this bound to comment on the possible detection of EP violation in different modified gravity scenarios. 

\subsection{Signal to noise for the bispectrum}

The signal to noise calculation closely follows the standard calculation for the case of primordial non-Gaussianities (see for example \cite{Scoccimarro:2003wn}). We will assume a survey of a given comoving volume $V$ which defines the fundamental scale in momentum space, $k_f = 2\pi / V^{1/3}$. In this setup, the bispectrum estimator is given by
\be
B(k_1,k_2,k_3) = \frac{V_f}{V_{123}} \int_{k_1} \mathrm d^3 q_1  \int_{k_2} \mathrm d^3 q_2  \int_{k_3} \mathrm d^3 q_3 \; \delta(\vec q_1 + \vec q_2 + \vec q_3) \cdot \delta_{\vec q_1}\delta_{\vec q_2}\delta_{\vec q_3} \;,
\ee
where $V_f=(2\pi)^3/V$ is the volume of the fundamental cell, the integration is done over the spherical shells with bins defined by $q_i \in (k_i-\delta k/2, k_i+\delta k/2)$ and
\be
V_{123} \equiv \int_{k_1} \mathrm d^3 q_1  \int_{k_2} \mathrm d^3 q_2  \int_{k_3} \mathrm d^3 q_3 \; \delta(\vec q_1 + \vec q_2 + \vec q_3) \approx 8\pi^2 \; k_1 k_2 k_3 \; \delta k^3 \;.
\ee
We will assume no significant correlation among different triangular configurations or, in other words, that the bispectrum covariance matrix  is diagonal and given by a Gaussian statistics. It can be shown that in this case the variance is given by \cite{Scoccimarro:2003wn}
\be
\label{bispectrum_variance}
\Delta B^2(k_1,k_2, k_3) = k_f^3 \frac{s_{123}}{V_{123}} P_{\text {tot}}(k_1) P_{\text {tot}}(k_2) P_{\text {tot}}(k_3) \;,
 \ee
where $s_{123}=6,2,1$ for equilateral, isosceles and general triangles, respectively. The power spectrum $P_{\text {tot}}(k)$ is given by
\be
P_{\text {tot}}(k) = P(k) + \frac{1}{(2\pi)^3} \frac{1}{\bar  n} \;,
\ee
where the last term on the right hand side  accounts for the shot noise and $\bar n$ is the number density of galaxies in the survey. In what follows we will neglect the shot noise contribution because we want to estimate the total amount of signal in principle available for a survey of a given volume, without restricting our analysis specifically to galaxy surveys. Moreover, for our estimates we will use only modes that are in the linear regime where the shot noise is expected to be negligible.

Given these definitions, the signal-to-noise ratio is calculated as
\be
\left( \frac SN \right)^2 = \sum_{T} \frac{\left( B_{\text{new physics}}(k_1,k_2,k_3) - B_{\text{standard}}(k_1,k_2,k_3)\right)^2}{\Delta B^2(k_1,k_2,k_3)} \;,
\ee
where the sum runs over all possible triangles formed by $\vec k_1$, $\vec k_2$ and $\vec k_3$ given $k_{\mathrm{min}}$ and $k_{\mathrm{max}}$. Typically, the sum is written down such that the same triangles are not counted twice and the symmetry factor $s_{123}$ takes care of special configurations. In our case, with two different species of particles, the bispectrum is not symmetric when momenta are exchanged and the previous equations have to be modified accordingly. We will impose $s_{123} = 1$ for all configurations and the sum over triangles will be
\be
\sum_{T} \, \equiv \sum_{k_1=k_{\rm min}}^{k_{\mathrm{max}}} \, \sum_{k_2=k_{\rm min}}^{k_{\mathrm{max}}} \, \sum_{k_3=k_{\rm min}^*}^{k_{\rm max}^*} \;\;,
\ee
where $k_{\rm min}^* \equiv \text{max}(k_{\rm min},|\vec k_1 - \vec k_2|)$, $k_{\rm max}^* \equiv \text{min}(|\vec k_1+ \vec k_2|, k_{\rm max})$ and the discrete sum is done with   $|\vec k_{\mathrm{max}}-\vec k_{\rm min}|/\delta k$ steps where $\delta k$ is a multiple of $k_f$. In the following we fix $\delta k = k_f$.

%To get an estimate on the constraint we could obtain on the parameters of the "new physics", we require that the signal to noise must be of order one to have detection.

\subsection{Estimate for our toy model}

Now that we have defined the estimator, we  apply it to the case of violation of the EP. We will not restrict ourselves to squeezed triangle configurations but we  exploit all possible triangular configurations of eq.~\eqref{3-p}.  

In the case at hand, the signal to noise  takes the form 
\be
\left( \frac SN \right)^2 = \sum_{T} \frac{\left[ B^{(AB)}_{ \alpha^2}(k_1,k_2,k_3) - B^{(AB)}_{\alpha^2=0}(k_1,k_2,k_3)\right]^2}{\Delta [B^{(AB)}]^2(k_1,k_2,k_3)} \;,
\ee
where the bispectrum $B^{( AB)}(k_1,k_2,k_3)$  is defined by
\be
\langle \delta_{\vec k_1} (\eta) \dA_{\vec k_2}(\eta) \dB_{\vec k_3}(\eta) \rangle = (2 \pi)^3 \delta_D(\vec k_1+ \vec k_2 + \vec k_3) B^{( AB)}(k_1,k_2,k_3) \;,
\ee
where the left hand side of this equation is computed from eq.~\eqref{3-p} at leading order in $\alpha^2$. For the computation we employ 
\be
\delta^{(A)}_{\vec k} (\eta) \equiv \frac{1}{w_A + w_B b} \delta_{\vec k} (\eta)\;, \qquad \delta^{(B)}_{\vec k} (\eta) \equiv \frac{1}{w_A/b + w_B } \delta_{\vec k} (\eta)\;, 
\ee
where we  compute the  transfer function of the total matter density contrast $\delta$ using the code CAMB \cite{Lewis:1999bs}. 
We then define 
\be
\langle \delta_{\vec k} (\eta) \delta_{\vec k'}(\eta) \rangle = (2 \pi)^3 \delta_D(\vec k+ \vec k' ) P(k) \;, \qquad \langle \delta^{(X)}_{\vec k} (\eta) \delta^{(X)}_{\vec k'}(\eta) \rangle = (2 \pi)^3 \delta_D(\vec k+ \vec k' ) P^{(X)}(k) \;.
\ee
Following eq.~\eqref{bispectrum_variance}, for the variance of the bispectrum we use
\be
\Delta \big[B^{( AB)}\big]^2(k_1,k_2, k_3) = k_f^3 \frac{s_{123}}{V_{123}} P(k_1) P^{(A)}(k_2) P^{(B)}(k_3) \;.
\ee

\begin{figure}
\centering
\includegraphics[width=3.25in]{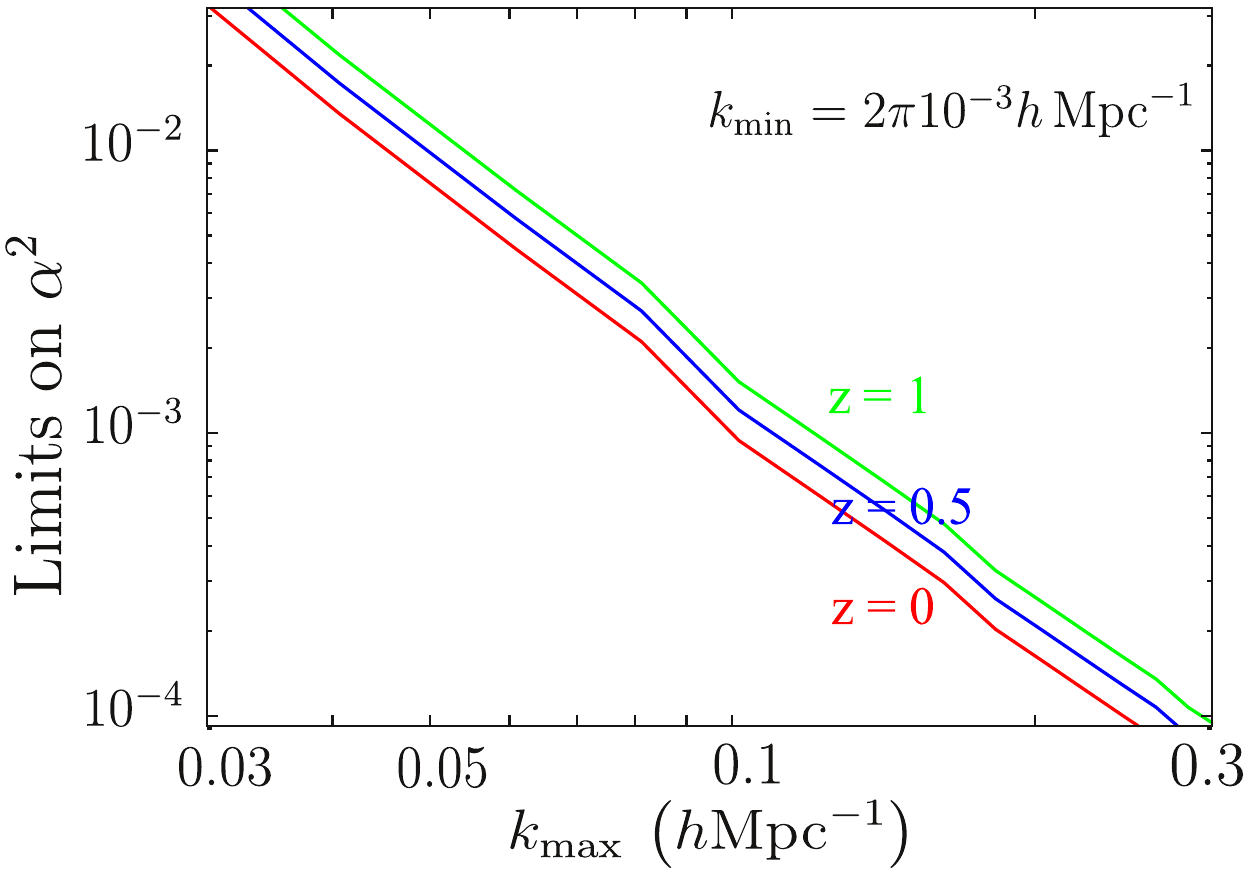} \hspace{0.2cm} \includegraphics[width=3.25in]{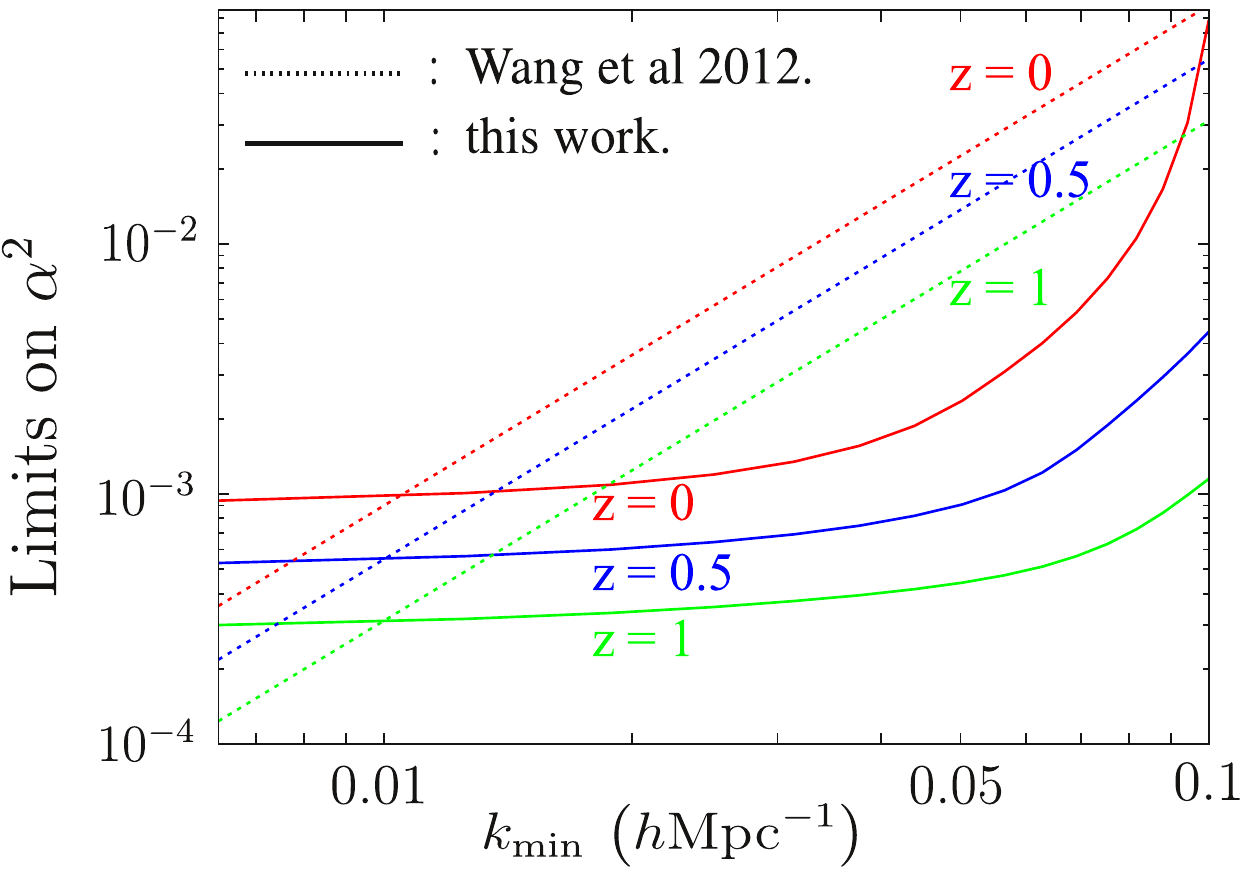} 
\caption{\small Expected error on $\alpha^2$, $\sigma (\alpha^2)$, for a survey with volume $V=1(\mathrm{Gpc}/h)^3$ at three different redshifts, $z=0$, $z=0.5$ and $z=1$. Left: $\sigma (\alpha^2)$ is plotted as a function of $k_{\mathrm{max}}$. We have chosen $k_{\rm min} = 2 \pi/V^{1/3}$ so that the violation of the EP extends to the whole survey. Right: $\sigma (\alpha^2)$ is plotted as a function of $k_{\mathrm{min}}$. $k_{\rm max}$ is given by $ 0.10, \, 0.14, \, 0.19$ for $z=0, \, 0.5, \, 1$ respectively. The dotted lines represent $\alpha^2 \lesssim 10^{-6} (m/H)^2$, i.e.~the bound on $\alpha^2$ from screening the Milky Way \cite{Wang:2012kj}.}
\label{figure1}
\end{figure}
Figure \ref{figure1} shows the estimated error on $\alpha^2$, $\sigma (\alpha^2)$, for three different  surveys of volume $V= 1 (h^{-1} \text{Gpc})^3$  at redshift $z=0$, $z=0.5$ and $z=1$, respectively.
On the left panel this is shown as a function of $k_{\rm max}$ for the
smallest possible $k_{\rm min}$, i.e.~$k_{\rm min} = k_f =2 \pi
/V^{1/3}$. The smallest measurable value of $\alpha^2$ roughly scales
as $k_{\rm max}^{-2.8}$, so that it crucially depends on our ability
to capture the shortest scales.\footnote{For $k_{\rm max} \sim
  0.1h^{-1} \text{Mpc}$ we roughly agree with \cite{Kehagias:2013rpa}
  but we find a different dependence on $k_{\rm max}$.
Our result can be roughly understood as follows. 
Recall that the experimental constraint on  $f^{\rm loc}_{\rm NL}$ goes like
$\Delta f^{\rm loc}_{\rm NL} \sim 5  \sqrt{10^6/ N}$, where $N$ is the number of
modes. This is consistent with the fact that the Planck limit is about $f^{\rm loc}_{\rm NL} \lesssim 5$, for $N \sim 10^6$ \cite{Ade:2013ydc}. For the large scale structure we
are interested in here, $N \sim (k_{\rm max}/k_{\rm min})^3 \sim 4 \cdot 10^3$, and so $\Delta f^{\rm loc}_{\rm NL} \sim
80$, which is consistent with Fig.~3 of Ref.~\cite{Scoccimarro:2003wn}. In the case of the EP violation we effectively have $f^{\rm loc}_{\rm NL} \sim \alpha^2 \times q \, k / (\Omega_m H_0^2)$.
Assuming $k/q \sim 10^2/(2 \pi)$, we therefore have a bound of $\alpha^2$
that is about $ 4 \cdot 10^{-3}$. This argument also tells us the
scaling with $k_{\rm max}$: $\sqrt{N} \propto k_{\rm max}^{3/2}$, 
and the scaling of the effective $f^{\rm loc}_{\rm NL}$  adds one more
power of $k_{\rm max}$, giving us a limit on $\alpha^2$ that scales
as $k_{\rm max}^{-2.5}$, roughly agreeing with our $k_{\rm max}^{-2.8}$ scaling.}

On the right panel, the estimated relative variance is shown as a function of  $k_{\rm min}$. For each  survey, we take $k_{\rm max}$ such that we are still in a quasi-linear regime where theoretical control in perturbation theory is  possible. In particular, we fix  $k_{\rm max}=\pi/(2R)$ where $R$ is chosen in such a way that $\sigma_{R} $, the root mean squared linear density fluctuation of the matter field in a ball of radius $R$, is $0.5$.(\footnote{Apart from the theoretical uncertainty in understanding the nonlinear regime of density fluctuations, other effects neglected here hinder the access to small scales. In redshift surveys, the smallest scales are  affected by the radial smearing due to redshift distortion that are uncorrelated with the density fluctuations, such as the one coming from the Doppler shift due to the virialized motion of galaxies within clusters or the one due to the redshift uncertainty of spectroscopic galaxy samples. See for instance \cite{Huang:2012mr} for a discussion.})    
This yields $k_{\rm max}= 0.10, \, 0.14, \, 0.19$ for $z=0, \, 0.5, \, 1$ respectively. From Fig.~\ref{figure1} we see that the dependence on $k_{\text{min}}$ is very mild when going to zero. This seems counterintuitive, because eq.~\eqref{VEP}  indicates that the bispectrum diverges as $1/q$ at small $q$, giving more signal. However, in that limit the power spectrum of matter fluctuations scales as $q$, $P(q) \propto q$,  canceling the enhancement. This differs from the familiar case of  local non-Gaussianity where  the divergence scales as $1/q^2$, causing the known increase of precision on $f_{\rm NL}^{\rm loc}$ when going to larger surveys. 
The improvement of the constraints at higher redshifts, discussed also in \cite{Kehagias:2013rpa}, is due to the fact that $k_{\rm max}$ increases and, assuming a fixed volume, we have access to more modes.
Our constraints can be compared with that for chameleon models derived in Ref.~\cite{Wang:2012kj} from requiring that the Milky Way must be screened. This yields
\be
\alpha^2 \lesssim 10^{-6} (m/H)^2\;,
\ee 
where $m$ is the Compton mass of the chameleon. In this case $k_{\rm min}$ can be identified with $m^{-1}$, the  Compton wavelength of the chameleon and one sees that for $m^{-1} \gtrsim 0.01$ our constraints can improve that of Ref.~\cite{Wang:2012kj}.

When looking for EP violation, a possible contaminant is the initial density or velocity bias between two different species. For instance, even in single-field inflation we know that baryons and dark matter have different initial conditions on scales below the sound horizon at recombination, because at recombination baryons are tightly coupled to photons through Thomson scattering,  while  dark matter particles are free falling.
As discussed in  \cite{Tseliakhovich:2010bj,Bernardeau:2011vy,Bernardeau:2012aq}, the  relative velocity between baryons and dark matter excites long wavelength isodensity modes that couple to small scales reducing the formation of early structures. 
However, one can check that this effect decays more rapidly than the one described by eq.~\eqref{eq:EPviolation}. For instance, assuming no violation of the EP but an initial  density and velocity bias between the two species $A$ and $B$, $u_a=(b_A,b_A,b_B,b_B)$, one obtains
\be
\label{eq:Hirata}
\lim_{q\to 0} \langle \delta_{\vec q}(\eta) \delta^{(A)}_{\vec k_1}(\eta) \delta^{(B)}_{\vec k_2}(\eta) \rangle' \simeq 4 \, (b_A-b_B)e^{-\frac32(y-y_0)} \, \frac{\vec{k}\cdot\vec{q}}{q^2} P(q,\eta) P_{AB}(k,\eta) \, ,
\ee
independently of $w_A$ and $w_B$. Thus, the effect is still divergent as $1/q$ but rapidly decays, so that it is typically suppressed by a factor $\sim (1+z_0)^{-3/2}$ where $z_0$ represents the initial redshift. 
For the example discussed above  of baryons and dark matter we can take $z_0 \simeq 1100$ and today this effect is thus suppressed by $\sim {\cal O}(10^{-5})$. Moreover, if we use galaxies  to probe the EP it will be further suppressed by the fact that the baryon-to-dark matter ratio is rather constant in different galaxies.

When using galaxies, one should also remember that their density field is a biased tracer and that in general we expect the bias to contain nonlinearities. Thus,  other contributions are expected in eq.~\eqref{VEP}, for instance of ${\cal O} [(k/q)^0]$ if the nonlinear bias is scale independent. 
To compute the signal-to-noise ratio correctly taking into account this effect, one should include these nonlinear contributions and marginalise over the bias parameters, similarly to what done in the context of non-Gaussianity, for instance in Ref.~\cite{Scoccimarro:2003wn}. However, due to its different scale and angular dependence, we do not expect the marginalization over nonlinear bias to dramatically change our estimates.

Before concluding, it is important to stress that our estimates so far assume that we know which are the two classes of objects that violate the EP. In practice, one will have to classify objects either based on some intrinsic property (mass, luminosity, dark matter content) or some environmental property (like being inside an overdense or underdense region), and astrophysical uncertainties in the selection of these objects may significantly suppress the signal. In particular, if the kind of objects we aim for is quite rare, the shot noise, which we have neglected so far, will be an important limitation. In this sense the limits discussed above are the most stringent one can get for an ideal survey of a given volume.

Despite these limitations, the absence of any signal when the EP holds is very robust. The long mode cannot give any $1/q$ effect, independently of the bias of the objects and of the selection strategy we use.
Furthermore, the same statement holds also in redshift space \cite{Creminelli:2013poa, Kehagias:2013rpa}, which makes the connection with observations even more straightforward. In other words, all the complications that enter when one wants to use the data to infer the underlying dark matter 3-point function are not relevant here if we only want to show that the EP is violated. Of course, once a violation is detected, it would be much more challenging to better characterize the source of the violation.

\section{\label{sec:models}Modifications of gravity and equivalence principle violation}
A violation of the consistency relations requires a macroscopic violation of the EP: different astrophysical objects must fall at a different rate. One can envisage various possibilities depending on which is the relevant feature that determines the EP violation.

{\bf Baryon content.} If dark matter and baryons have a different coupling with a light scalar, one has a violation of the EP at the fundamental level. This causes different astrophysical objects, with a different baryon/dark matter ratio to fall at a different rate in an external field. This scenario is however very constrained: Planck \cite{Pettorino:2013oxa} limits this kind of couplings to be $\lesssim 10^{-4}$ smaller than gravity. This is far from what we can achieve with our method, since most astrophysical objects have a quite similar baryon content and this suppresses substantially the EP violation.

{\bf Amount of screening.} The screening of extra forces to satisfy the gravity tests in the solar system induces violations of the EP \cite{Hui:2009kc}. We can distinguish various cases, depending on the screening mechanism.

For {\em chameleon} \cite{Khoury:2003aq} or {\em symmetron} \cite{Hinterbichler:2010es,Pietroni:2005pv,Olive:2007aj} screening the EP violation can be of order unity between screened and unscreened objects. However, the necessity of screening inside the solar system limits the impact of the fifth force on cosmological scales. Indeed, one can find a model-independent limit on the mass of the scalar \cite{Wang:2012kj,Brax:2011aw}
\be
m^2 \gtrsim 10^6 \alpha^2 H^2\;.
\ee  
This inequality, which is valid at low redshifts, limits the effect of the scalar on short scales $k/a \lesssim m$. In Fig.~\ref{figure1} we compare this limit with our signal to noise forecast at different redshifts: a detection of EP violation is possible, though quite challenging. The screening here depends on the typical value of the gravitational potential $G M/r$ of the object. Given that we know the Milky Way is screened, one should look for objects with a lower $\Phi$ to find unscreened objects. This looks challenging since in a survey one is typically sensitive to galaxies which are more luminous and therefore more massive than the Milky Way.

For {\em Galileon} screening \cite{NRT2009} the issue of EP violation is rather subtle. On one hand, one can show that an object immersed in an external field which is constant over the size of the object will receive an acceleration proportional to the mass and independent of the possible Vainshtein screening of the object \cite{Hui:2009kc}. On the other hand, given the nonlinearity of the scalar equations, the value of the external field may not be the same before and after the object is put into place. For example, the Moon changes the solution of the Galileon around the Earth and the nonlinearity of the system is such that the acceleration the Moon experiences is different from the one of a test particle orbiting around the Earth \cite{Hiramatsu:2012xj,Belikov:2012xp}. This complicated nonlinear behaviour is difficult to control in general, but we can however prove that the Galileon models do {\em not} lead to violations of our consistency relations. Well inside the horizon, structure formation in the presence of the Galileon $\pi$ follows the equations
\begin{align}
\dot{\vec v} +(\vec v \cdot \vec \nabla) \vec v  & = - \vec\nabla \Phi - \alpha\vec\nabla \pi\;, \\
F(\partial_i\partial_j\pi) & = \alpha 8 \pi G \rho \label{eq:Galeq} \;, \\
\nabla^2 \Phi & = 4 \pi G \rho \;,
\end{align}
where $F$ is the equation of motion for the Galileon, which only
depends on the second derivatives of $\pi$. The point is that one can
run the same argument as in the absence of the Galileon: a homogeneous
$\vec\nabla(\Phi + \alpha \pi)$ can be removed by a change of
coordinates that brings us to an accelerated frame \footnote{Notice
  that we can remove a homogeneous field, with arbitrary
  time-dependence. This is not a symmetry of the full Galileon theory,
  but it holds deep inside the Hubble radius, when time-derivatives
  can be neglected in eq.~\eqref{eq:Galeq}.}. For this to happen the
symmetry of Galileons is crucial, since it makes a homogeneous
gradient of $\pi$ drop out of eq.~\eqref{eq:Galeq} \cite{HN2012}. 
(This does not work, for example, in the case of the chameleon.) The
homogenous gradient can describe a long mode in the linear regime
(simulations
\cite{Cardoso:2007xc,Khoury:2009tk,Schmidt:2009sg,Chan:2009ew,Schmidt:2009sv}
show that the scalar force is active, i.e.~not Vainshtein suppressed,
on sufficiently large scales) so that, barring primordial
non-Gaussianity, the effect of a long mode boils down to the change of
coordinates, which does not give any effect at equal time.
\footnote{The reader might wonder how one can reconcile the lack of consistency
  relation violation, with the known equivalence principle violation
(at a small level) in the case of the Galileon. The point is that
the boundary condition in the Earth-Moon example is quite different
from that in the cosmological example. In cosmology, we know from numerical
simulations that $\pi$ is in the linear regime on large scales; in the
Earth-Moon example, it is a computation entirely within the Vainshtein
radius of the system.
}

An intermediate case between the ones above is given by {\em K-mouflage} \cite{Babichev:2009ee}, where the screening depends neither on the value of the field---like in the chameleon---nor on the value of the second derivatives---like in the Galileon---but on the first derivative. This happens when we have a generic kinetic term of the form $P(X)$ with $X \equiv (\partial\phi)^2$. Although this case has not been thoroughly studied, there is no reason to expect our consistency relations to work since, in the absence of Galileon symmetry, the argument above does not go through. In this case the screening depends on the typical value of $\nabla\Phi$ of the object.

{\bf Gravitational potential.} The no-hair theorem implies that black-holes do not couple with a scalar force. More generally, the mass due to self-gravity will violate the EP in the presence of a fifth force. Unfortunately, it seems impossible to observe isolated objects with a sizable component of gravitational mass. The mass of clusters only receives a contribution in the range $10^{-5} \div 10^{-4}$ from the gravitational potential and the correction is even smaller for less massive objects. Black holes, whose mass is entirely gravitational in origin, do not significantly contribute to the mass of the host galaxy.

{\bf Environment.} Another possibility is to divide the objects depending not on some intrinsic feature but on their environment, for example comparing galaxies in a generic place against galaxies in voids \cite{Kehagias:2013rpa}. The fifth force tends to be screened in a dense environment (blanket screening), while it is active in voids. Notice that this is not a test of the Galilean EP (different objects fall at the same rate in the same external field), but it still checks whether the effect of the long mode can be reabsorbed completely by a change of coordinates. The arguments made above for the Galileon case work also here and we expect no violation of the consistency relation in this case. This effect will be present in the case of chameleon screening (with the same limitations on the Compton wavelength discussed above) and in K-mouflage.

\section{Conclusions}
In this paper we discussed a method to test the Equivalence Principle on cosmological scales based on the recently proposed consistency relations for Large Scale Structure. The idea is simply that a homogeneous gravitational potential can be {\em exactly} removed by a suitable change of coordinates \cite{Creminelli:2013mca}. This is not true if the EP is violated, in which case $\epsilon \neq 0$ in eq.~\eqref{VEP}.

The method that we propose is very robust because the absence of a $1/q$ signal when the EP holds is not affected by nonlinearities at short scales, baryon physics, the issue of bias, redshift-space distortions and the way objects are selected \cite{Creminelli:2013mca,Creminelli:2013poa,Kehagias:2013yd,Peloso:2013zw,Kehagias:2013rpa}. 
Moreover, the signal of EP violation in the 3-point function cannot be confused with one due to primordial non-Gaussianity. The reason is that, due to the parity of the 2-point function, the squeezed limit of the primordial 3-point function cannot have a dipolar structure of the form \eqref{VEP}. Indeed, there are models of inflation which induce $1/q$ dependence of the 3-point function in the squeezed limit, such as Quasi-Single Field \cite{Chen:2009zp} or Khronon  Inflation \cite{Creminelli:2012xb}, but in these cases the 3-point function in the squeezed limit is a function of $q$ only and not of its direction. In models where the 3-point function in the squeezed limit  depends on the direction of $\vec q$, such as Solid Inflation \cite{Endlich:2012pz}, this dependence has a quadrupolar structure.

In conclusion, assuming there is no primordial non-Gaussianity, any appearance of  $1/q$ divergences in the 3-point function would be a clear signal of  violation of the EP.  Therefore, even though most of the models that violate the EP are either very constrained or produce small effects, the proposed test is so general that it deserves to be done. One can even take an agnostic point of view and, without referring to any particular model, try to explore correlations among different types of objects in $N$-body simulations  or directly in the data. For instance, as explained above, one aspect that has not been studied in the literature is EP violation in scalar-tensor theories with a generic kinetic term $P(X)$ \cite{Babichev:2009ee}.
It would be interesting to analyze the screening in these theories and directly observe violations of the type of eq.~\eqref{VEP} in $N$-body simulations. Testing the EP in the data using our method   will become particularly relevant for  forthcoming large scale structure surveys, whose volumes will be large enough to put interesting limits on the violation of the EP on cosmological scales. In this case, one needs to go beyond what done in \cite{Creminelli:2013mca,Creminelli:2013poa} and carefully include relativistic effects in galaxy surveys and a treatment of redshift-space distortions beyond the plane-parallel approximation. We leave this for the future.

Notice that the same limit of the 3-point function of eq.~\eqref{VEP}, when the long mode is taken outside the Hubble scale, induces a dipolar modulation of the cross power spectrum between objects $A$ and $B$. The modulation is of order\footnote{The $1/q$ behavior is only valid in the non-relativistic limit and it saturates at the Hubble scale, see \cite{Creminelli:2013mca}.}
\be
\label{anisotropy}
\epsilon \, \Phi_{\rm L} \frac{\vec k \cdot \hat q}{H} \;,
\ee
where the direction $\hat q$ is fixed by the average over long modes.  Although the anisotropy is suppressed by the long-mode amplitude, it grows going to short scales where it can become significant. Limits on the anisotropy of the auto-power spectrum are presently of order ${\cal O}(10^{-2})$ \cite{Hirata:2009ar} and it would be interesting to see what can be done using different objects, although it is difficult that this will do better than  directly looking at the 3-point function.

\subsection*{Acknowledgements}
It is a pleasure to thank D.~Baumann, P.~Brax, B.~Horn, J.~Khoury, D.~L\'opez
Nacir, M.~Pietroni, R.~Scoccimarro, G.~Trevisan, L.~Verde and especially E.~Sefusatti
for very useful discussions. J.G. and F.V. acknowledge partial support by
the ANR {\it Chaire d'excellence} CMBsecond ANR-09-CEXC-004-01.
LH acknowledges support by the US DOE under 
grant DE-FG02-92-ER40699 and NASA under ATP grant NNX10AN14G.

\end{document}